# CMB Anisotropies in the Weak Coupling Limit

Wayne Hu & Martin White

Departments of Astronomy and Physics and Center for Particle Astrophysics,
University of California, Berkeley, CA 94720-7304, USA

**Abstract.** We present a new, more powerful and accurate, analytic treatment of cosmic microwave background (CMB) anisotropies in the weakly coupled regime. Three applications are presented: gravitational redshifts in a time dependent potential, the Doppler effect in reionized scenarios, and the Vishniac effect. The Vishniac effect can dominate primary anisotropies at small angles even in late and minimally reionized models in flat dark-matter dominated universes with Harrison-Zel'dovich initial conditions. The techniques developed here refine previous calculations yielding a larger coherence angle for the Vishniac effect and moreover can be applied to non-trivial ionization histories. These analytic expressions may be used to modify results for the standard cold dark matter model to its cosmological constant and reionized extensions without detailed and time consuming recalculation.

## 1. Introduction

Damping processes in the limit that last scattering of cosmic microwave background (CMB) photons lasted only a short duration have been the subject of much recent interest (see *e.g.* Hu & Sugiyama 1995a; Bond 1995) and require detailed information on the microphysical interactions that govern photon-baryon coupling (see *e.g.* Hu, Scott, Sugiyama, & Silk 1995). Here we focus on the opposite limit of a very gradual decoupling of the photons from interactions with the matter. This occurs for scattering effects in reionized models and gravitational effects between last scattering and the present. We significantly improve previous analytic treatments of this case (Kaiser 1984, Efstathiou 1988, Hu & Sugiyama 1994) by calculating the anisotropy directly rather than inferring it from the corresponding inhomogeneity of the CMB at last scattering.

These more accurate calculational techniques are especially important in the case of higher order effects which can dominate on scales under the thickness of the last scattering surface. Here linear contributions have been damped away and the anisotropy cannot be obtained by the standard procedure of numerically solving the *linearized* Boltzmann equation (Wilson & Silk 1982, Bond & Efstathiou 1984, Vittorio & Silk 1984). The dominant term is the second order Doppler contribution called the Vishniac effect (Ostriker & Vishniac 1986; Vishniac 1987). Because it is second order, Vishniac contributions are strongly weighted toward the present (Hu, Scott, & Silk 1994) and cannot be considered as a projection of temperature inhomogeneities *at* the last scattering surface. Indeed, 50% of the signal in temperature fluctuations for a reionized cold dark matter (CDM) model comes from redshifts of $z < 5$. This approaches the redshift at which the universe must be reionized to satisfy Gunn-Peterson constraints. Thus even minimally ionized CDM models can be expected to have arcminute contributions to temperature fluctuations from the Vishniac effect. The standard formalism (Efstathiou & Bond 1987; Efstathiou 1988) cannot adequately treat this important case in which last scattering does not occur on a single well defined surface. We evaluate the amplitude and detailed shape of the resultant minimal Vishniac anisotropy. Although it may well be the dominant small scale effect, its presence is unlikely to confuse measurements of the sharp damping tail of linear contributions. This has important implications for the measurement of the geometry and/or thermal history of the universe from small scale anisotropies.

To verify these new analytic techniques, we first develop them on and apply them to first order contributions where comparisons with highly accurate numerical treatments are possible. We present an analysis of first order Doppler effects on small scales which improves upon previous analytic work (Kaiser 1984) especially in the late ionization limit. The weak coupling approximation also accurately describes even the large angle effects of gravitational potential decay in a cosmological constant dominated universe. The resultant analytic formulae greatly simplify previous treatments (Kofman & Starobinski 1985) without compromising accuracy and allow for a quick and efficient calculation of this effect.

## 2. First Order Effects

Let us first verify the weak coupling approximation in the case of linear contributions which can be calculated precisely by numerical methods. To first order in density fluctuations, the Boltzmann equation in Fourier space for the evolution of photon temperature perturbations $\Delta T/T = \Theta(\mathbf{k}, \boldsymbol{\gamma}, \eta)$ in a flat $\Omega_0 + \Omega_\Lambda$ universe has the formal solution

$$[\Theta + \Psi](\mathbf{k}, \boldsymbol{\gamma}, \eta) = \int_0^\eta [(\Theta_0 + \Psi + \boldsymbol{\gamma} \cdot \mathbf{v}_b)\dot{\tau} + 2\dot{\Psi}] \quad (1)$$
$$\times e^{-\tau} e^{ik\mu(\eta'-\eta)} d\eta',$$

[see Hu & Sugiyama 1995b, eqn. (54)] where we have neglected the small corrections from the angular dependence of Thomson scattering and polarization. Here $\Theta_0$ is the isotropic temperature fluctuation, $\Psi$ is the Newtonian gravitational potential, $\mathbf{v}_b$ is the baryon velocity with $c = 1$, $\mathbf{k} \cdot \boldsymbol{\gamma} = k\mu$, and overdots represent derivatives with respect to conformal time $\eta = \int dt/a$ where $a$ is the scale factor normalized to the present. The differential optical depth to Compton scattering is given explicitly by $\dot{\tau} = x_e n_e \sigma_T a$, where $x_e$ is the ionization fraction, $n_e$ is the electron number density, and $\sigma_T$ is the Thomson cross section. From left to right, the linear contributions to the anisotropy in the integrand are the intrinsic photon temperature at last scattering (Peebles & Yu 1970), the gravitational redshift (Sachs & Wolfe 1967), the Doppler effect (Sunyaev & Zel'dovich 1970), and the differential gravitational redshift or integrated Sachs-Wolfe (ISW) effect (Sachs & Wolfe 1967). The first three effects contribute at last scattering. Consequently, these sources are multiplied by the differential visibility function $\dot{\tau} e^{-\tau}$ which is the probability of last scattering in the interval $d\eta$ at $\eta$. The ISW effect contributes between last scattering and the present as the $e^{-\tau}$ suppression at high optical depth shows.

Below the horizon but above the photon diffusion scale at last scattering, the photons and baryons evolve adiabatically leading to acoustic oscillations as photon pressure resists gravitational compression (Peebles & Yu 1970). This leads to strong contributions from the intrinsic photon temperature and the "Doppler peak" structure of anisotropies from standard recombination at $z_* \approx 1000$ (see e.g. Bond & Efstathiou 1987; Doroshkevich, Zel'dovich & Sunyaev 1978). However if the universe is significantly reionized, last scattering occurs much later. In this case, the Compton optical depth $\tau$ drops below unity only after the free electron density has been sufficiently decreased by the expansion $z_* \approx 100(\Omega_0 h^2/0.25)^{1/3}(x_e \Omega_b h^2/0.0125)^{-2/3}$. At this time, the diffusion length approaches the horizon scale. Since diffusion damps out intrinsic temperature fluctuations in the CMB as $e^{-\tau}$, acoustic oscillations do not appear in the present day anisotropy of strongly reionized models.

In the reionized case, the other sources in equation (1), which arise from fluctuations in the matter, are important. These sources can yield strong contributions if last scattering occurs after the matter has been released from Compton drag. After $z_d \approx 160(\Omega_0 h^2)^{1/5} x_e^{-2/5}$, perturbations grow as in linear theory. Here the Hubble constant is given by $H_0 = 100h$ km s$^{-1}$ Mpc$^{-1}$. Contributions well after $z_d$ can be analytically calculated, following Kaiser (1984), by iteratively solving for the feedback of the Doppler term into the intrinsic temperature fluctuation at last scattering. Averaging equation (1) over angles and keeping the leading order term, we obtain for the first order effect

$$\Theta_0 + \Psi \approx \int_0^\eta \dot{\tau} e^{-\tau} iv_b j_1[k(\eta - \eta')] d\eta' \approx iv_b \dot{\tau}/k, \quad (2)$$

for scales below the horizon at $\eta$, $k\eta \gg 1$. Here diffusion and cancellation suppresses the contribution by $\dot{\tau}/k$, the optical depth through a wavelength. The $\ell$th Legendre moment of the temperature fluctuation $\Theta_\ell = i^\ell \frac{1}{2} \int_{-1}^1 \Theta P_\ell(\mu) d\mu$ from the secondary sources is given by

$$\Theta_\ell(k, \eta_0) = \frac{1}{k} \int_{\eta_d}^{\eta_0} [-i\dot{v}_b \dot{\tau} - iv_b \ddot{\tau} + 2k\dot{\Psi}] \quad (3)$$
$$\times e^{-\tau} j_\ell[k(\eta_0 - \eta)] d\eta, \qquad (\ell \geq 2)$$

which is valid for the Doppler term only under the horizon scale at last scattering.

At sufficiently small scales, the photon traverses many wavelengths of the fluctuation during the interval the sources in the integral of equation (3) contribute. In this case, opposing contributions from crests and troughs tend to cancel and damp the anisotropy. Mathematically, this can be seen since the spherical Bessel function in equation (3) oscillates rapidly in comparison with the source terms. Taking the slowly varying quantities out of the integral, noting that

$$\int_0^\infty j_\ell(x) dx = \frac{\sqrt{\pi}}{2} \frac{\Gamma[(\ell+1)/2]}{\Gamma[(\ell+2)/2]}, \quad (4)$$

and expressing the anisotropy in the conventional form of

$$C_\ell = \frac{2}{\pi} \int_0^\infty \frac{dk}{k} k^3 |\Theta_\ell(k, \eta_0)|^2, \quad (5)$$

we obtain

$$C_\ell^{(1)} = \frac{1}{2} \left( \frac{\Gamma[(\ell+1)/2]}{\Gamma[(\ell+2)/2]} \right)^2 \int \frac{dk}{k} \frac{k^3}{(k\eta_0)^6} \quad (6)$$
$$\times [S_D(\eta_\ell) + S_{\text{ISW}}(\eta_\ell)]^2 P(k),$$

where the matter power spectrum $P(k) = |\Delta_b(k, \eta_0)|^2$ and the Doppler and ISW sources,

$$S_D(\eta) = \left( \ddot{D}\dot{\tau} + \dot{D}\ddot{\tau} \right) \eta_0^3 e^{-\tau}, \quad (7)$$

$$S_{\text{ISW}}(\eta) = 3H_0^2 \Omega_0 \frac{D}{a} \left( \frac{\dot{D}}{D} - \frac{\dot{a}}{a} \right) \eta_0^3 e^{-\tau},$$

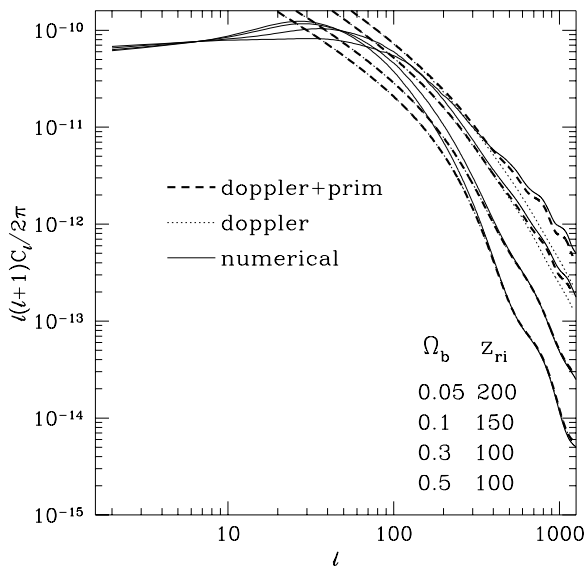

**Fig. 1.** Doppler effect in reionized CDM models ($\Omega_0 = 1$, $h = 0.5$, $n = 1$). For late ionization models $z_{ri} \lesssim z_d$, the analytic estimate which involves the neglect of Compton drag is an excellent approximation to the numerical results at small scales. For low $\Omega_b h^2$ models, the optical depth between $z_{ri}$ at which the universe becomes fully ionized and the present is insufficient to erase completely the primary signal from standard recombination. The total effect is thus described by adding $e^{-2\tau}$ of the standard recombination result to the Doppler effect. Here and in Figs. 3,4,5 the relative normalization is fixed by the amplitude of the matter power spectrum.

are evaluated at the peak of the $\ell$th Bessel function $\eta_\ell = \eta_0 - (\ell + 1/2)/k$. Here we assume linear theory growth for the density perturbations, as is appropriate after the drag epoch, $\Delta_b(k, \eta) = D(\eta)\Delta_b(k, \eta_0)$ and we have employed the continuity and Poisson equations to relate this to the velocities and potentials

$$k\mathbf{v}_b(\mathbf{k}, \eta) = \dot{\Delta}_b\hat{\mathbf{k}} = i\dot{D}\Delta_b(k, \eta_0)\hat{\mathbf{k}}, \quad (8)$$

$$k^2\Psi(k, \eta) = -\frac{3}{2}(D/a)H_0^2\Omega_0\Delta_b(k, \eta_0),$$

assuming the baryons follow the total matter. Note that in an $\Omega_0 = 1$ universe $D = (\eta/\eta_0)^2$. More generally, it is given by $D \propto H \int da/(aH)^3$, where $H$ is the time dependent Hubble parameter.

The Doppler and ISW contributions have the same angular scaling despite the fact that velocities dominate over potentials at small scales (high $k$) from equation (8) (Hu & Sugiyama 1994). This is because the Doppler contribution suffers severe crest-trough cancellation due to the irrotational nature of linear theory flows (Ostriker & Vishniac 1986). Crest-trough cancellation does not occur if the plane wave oscillates in a direction perpendicular to the line of sight $\mathbf{k} \perp \boldsymbol{\gamma}$. However, since $\mathbf{v}_b \parallel \mathbf{k}$ and the Doppler effect arises from the velocity component *along* the line of sight, it cannot escape cancellation through this mode. Thus ordinarily negligible effects may dominate small scale anisotropies.

In Fig. 1, we compare the analytic prediction for the Doppler effect with full numerical results. For late ionized low $\Omega_b h^2$ models, the optical depth between the reionization epoch and the present is insufficient to damp completely the primary signal from standard recombination. To describe the full effect at small angles one adds this contribution in quadrature with equation (6): $C_\ell^{(\mathrm{tot})} = C_\ell^{(1)} + e^{-2\tau}C_\ell^{(\mathrm{prim})}$. This simple expression may be useful for post-processing standard recombination results to account for reionization without detailed recalculation.

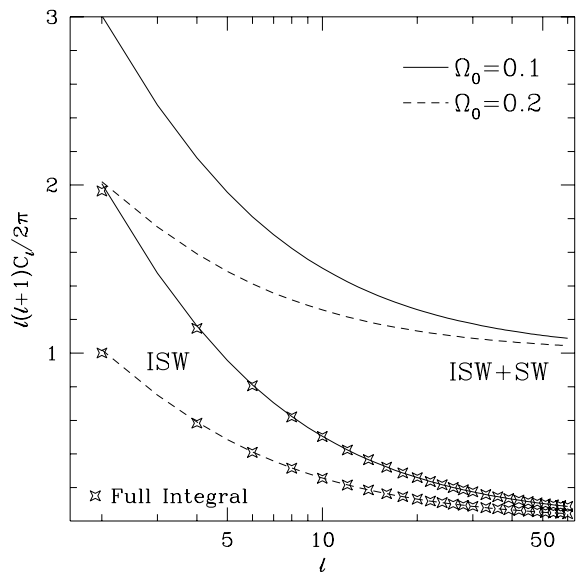

**Fig. 2.** Weak coupling approximation for the $\Lambda$ ISW effect. The analytic approximation based on crest-trough cancellation traces the full ISW integral to high accuracy and accounts for the falling ISW spectrum. As $\Omega_0$ increases, the relative contribution of the ISW effect decreases in comparison to the Sachs-Wolfe (SW) effect. Here we have chosen a pure power law $P(k) \propto k$ power spectrum.

Note that if reionization occurs sufficiently before the drag redshift and $\Omega_b h^2$ is low such that $z_* \sim z_d$, the baryons will not have completely fallen into the cold dark matter wells by last scattering. In this case, equation (6) overestimates the amplitude but not the angular dependence of the Doppler effect. Since previous analytic treatments (*e.g.* Kaiser 1984, Hu *et al.* 1994) also neglect Compton drag, we have improved the techniques for all cases.

On the other hand, if $\Omega_0 < 1$ the ISW term contributes at $\Lambda$-domination, $1 + z_\Lambda = (\Omega_\Lambda/\Omega_0)^{1/3}$, and the neglect of drag is an excellent approximation for all reasonable $\Omega_0$.

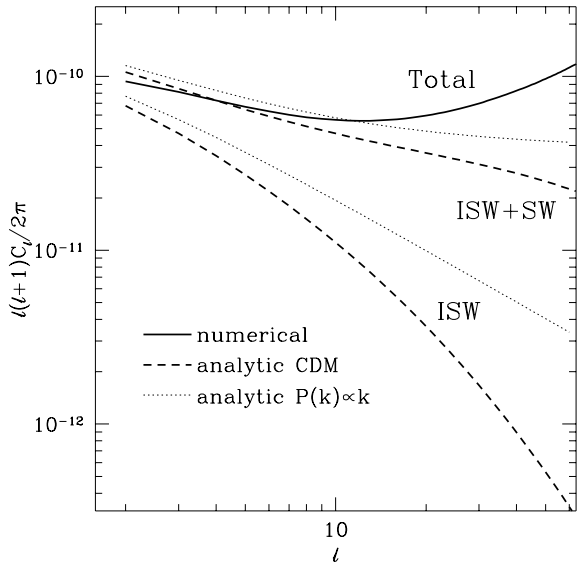

**Fig. 3.** Analytic versus numerical results for $\Lambda$ CDM. Comparison with numerical integration (thick solid lines) shows that at large scales, where $\Lambda$ ISW and Sachs-Wolfe (SW) contributions dominate the anisotropy, analytic formulae (thick dashed lines) adequately describe the total anisotropy. At small scales, the curves depart from the prediction due to gravitational redshifts from the radiation dominated era and acoustic contributions. Here we have chosen a high $h = 1.0$ to minimize these effects. Also plotted in thin dotted lines is the analytic prediction for a pure $P(k) \propto k$ power spectrum with the same normalization at large scales.

Furthermore, since the horizon at $\Lambda$ domination is close to the present horizon in size, the cancellation approximation is valid for all $k$'s relevant to anisotropies. Thus the above approximation provides a simple way of calculating the ISW contributions even at *large* angles for flat $\Lambda$ models. The combined Sachs-Wolfe and ISW effects can be written as $C_\ell = C_\ell^{(\mathrm{SW})} + C_\ell^{(\mathrm{ISW})}$, where

$$C_\ell^{(\mathrm{SW})} = \frac{2}{\pi} \int \frac{dk}{k} k^3 \left(\frac{1}{3}|\Psi(k,\eta_*)|\right)^2 j_\ell^2[k(\eta_0 - \eta_*)] \quad (9)$$

$$C_\ell^{(\mathrm{ISW})} = \frac{1}{2}\left(\frac{\Gamma[(\ell+1)/2]}{\Gamma[(\ell+2)/2]}\right)^2 \int \frac{dk}{k}\frac{k^3 P(k)}{(k\eta_0)^6} S_{\mathrm{ISW}}^2(\eta\ell).$$

In Fig. 2, we show that equation (9) is an excellent approximation of the full integral in equation (3).

For simple pure power law spectra $P(k) = Ak^n$, the Sachs-Wolfe integral can be evaluated analytically and the ISW term can be well approximated by a closed form expression. In particular, for the $n = 1$ case the combined effect becomes

$$\frac{\ell(\ell+1)C_\ell}{2\pi} \approx \frac{A}{8\pi^2} H_0^4 \Omega_0^{1.54}\left[1 + 78.5\left(\frac{\Gamma[(\ell+1)/2]}{\Gamma[(\ell+2)/2]}\right)^2 \quad (10)\right.$$

$$\left. \times (1 + 0.08/\ell)^{-1}(1 - \Omega_0^{0.094})^{2.07}\right].$$

where we have approximated $(\Omega_0 D_*/a_*)^2 \approx \Omega_0^{1.54}$ and the first and second terms in square brackets represent the Sachs-Wolfe and ISW effects respectively. For tilts up to $\pm 0.3$ around $n = 1$, the ISW integral is well fit by multiplying the $n = 1$ result by $[(\ell + 1/2)/\ell_p]^{n-1}$, where the pivot multipole is $\ell_p = 740\Omega_0^{-0.43}h^{-1}$. The Sachs-Wolfe result can be tilted by employing the standard gamma function formula (see *e.g.* White, Scott, & Silk 1994, eq. 15).

Realistic models are somewhat more complicated if high accuracy is required. In the $\Lambda$ CDM model of Fig. 3, processing of the initial $P(k) \propto k$ spectrum during radiation domination causes the present power spectrum to turn down at small scales. The error incurred by employing a power law approximation to the spectrum instead of the true processed spectrum is small at the quadrupole, but can be significant for somewhat smaller angles. Notice that the discrepancy is larger for the ISW effect than the Sachs-Wolfe effect since it is generated at later times when small scales subtend larger angles on the sky. This clearly shows the benefit of calculating anisotropies directly instead of via the spatial distribution where the differences in projection would be lost. We shall encounter this distinction again when we calculate the Vishniac effect in the next section.

## 3. The Vishniac Effect

Since the first order Doppler effect is mainly cancelled due to geometrical reasons associated with irrotational flows, second order effects can dominate the anisotropy at small angles. While in principle there are many sources in second order perturbation theory, it has been shown that the Vishniac effect is the dominant contribution wherever such effects are important (Hu et al. 1994, Dodelson & Jubas 1995). It is obtained by noting that fluctuations in the baryon density $\Delta_b$ change the optical depth $\tau$ to Compton scattering and hence the probability of scattering. The coupling of the optical depth fluctuation to the standard Doppler source yields the Vishniac term: $\gamma \cdot \mathbf{q}(\mathbf{x}) \equiv \gamma \cdot \mathbf{v}_b(\mathbf{x})\Delta_b(\mathbf{x})$.

Generalizing the Doppler term in equation (1), the formal solution to the Vishniac effect is

$$\Theta(\mathbf{k},\gamma,\eta) = \int_0^\eta \dot\tau e^{-\tau} \gamma \cdot \mathbf{q}\, e^{ik\mu(\eta'-\eta)} d\eta'. \quad (11)$$

We have neglected feedback effects from the generation of a monopole and quadrupole *by* the Vishniac effect into the full Boltzmann equation. For scales under the width of the visibility function $\dot\tau e^{-\tau}$, the Vishniac effect feeds back to create a monopole $\Theta_0 + \Psi = \mathcal{O}(q\dot\tau/k)$. Just as for the first order effect, these contributions are suppressed by the optical depth through a wavelength on small scales. For scales much larger than the width, $\boldsymbol{\gamma} \cdot \mathbf{q}$ sources only a dipole. Thus feedback effects are always negligible.

Following Vishniac (1987), let us decompose the solution at the present epoch $\eta_0$ into spherical harmonics,

$$\Theta(\mathbf{k}, \boldsymbol{\gamma}, \eta_0) = \sum_{\ell,m} \Theta_{\ell m}(\mathbf{k}) Y_{\ell m}(\Omega), \qquad (12)$$

so that

$$|\Theta_{\ell m}|^2 = \left| \int d\Omega\, Y_{\ell m}(\Omega) \int_0^{\eta_0} d\eta\, \dot{\tau} e^{-\tau} \boldsymbol{\gamma} \cdot \mathbf{q}\, e^{ik\mu(\eta-\eta_0)} \right|^2. \quad (13)$$

Choosing $\hat{\mathbf{z}} \parallel \mathbf{k}$, we note that the azimuthal angle dependence separates out components of $\mathbf{q}$ parallel and perpendicular to $\mathbf{k}$: $\boldsymbol{\gamma} \cdot \mathbf{q} = \cos\phi\sin\theta q_\perp + \cos\theta q_\parallel$. This dependence guarantees that the cross terms between the two components vanish after integrating over azimuthal angles. Thus the two contributions add in quadrature and may be considered as separate effects.

Let us consider the parallel component $\cos\theta q_\parallel = \mu q_\parallel$. This is to be integrated over the oscillatory function $e^{ik\mu(\eta-\eta_0)}$. If the wavelength is much smaller than the thickness of the visibility function, the integral will suffer severe cancellation unless $\mu = 0$. Just as in the first order case, unless the perturbation is perpendicular to the line of sight, we will be looking through many crests and troughs of the perturbation. Since the parallel component is proportional to $\mu$, there is no contribution from this direction, leaving a negligible total effect. Because the second order parallel term suffers the same cancellation as the first order term, it can always be neglected in comparison.

The perpendicular term does not vanish for $\mu = 0$ and thus survives cancellation. Since the final result after summing over all $\mathbf{k}$ modes has no preferred direction, define

$$|\Theta_\ell(k,\eta_0)|^2 \equiv \frac{1}{4\pi} \frac{1}{2\ell+1} \sum_m |\Theta_{\ell m}|^2, \qquad (14)$$

which corresponds to the former definition of $\Theta_\ell$. Using the addition theorem for spherical harmonics and the orthogonality of $\cos m\phi$, we find (Vishniac 1987)

$$|\Theta_\ell(k,\eta_0)|^2 = \frac{1}{8\ell(\ell+1)} \left| \int_{-1}^1 d\mu\, P_\ell^1 [(1-\mu^2)]^{1/2} \right. \qquad (15)$$
$$\left. \times \int_0^{\eta_0} d\eta\, \dot{\tau} e^{-\tau} q_\perp e^{ik\mu(\eta-\eta_0)} \right|^2.$$

The $\mu$ integral can be performed (Abromowitz & Stegun 1964, eqs [8.5.1, 10.1.14]), to yield

$$|\Theta_\ell(k,\eta_0)|^2 = \frac{1}{2}\ell(\ell+1) \left| \int_0^{\eta_0} d\eta\, \dot{\tau} e^{-\tau} q_\perp \frac{j_\ell(k\Delta\eta)}{k\Delta\eta} \right|^2, \quad (16)$$

where $\Delta\eta = \eta_0 - \eta$. Notice that this has a simple physical interpretation. We know from the spherical decomposition that a plane wave perturbation projects onto the shell at distance $\Delta\eta$ as $j_\ell(k\Delta\eta)$. The projection from $k$ to $\ell$ is not one-to-one and thus leads to an aliasing which is described by the oscillating tail of the Bessel function. If the amplitude of the plane wave is modulated by an angular dependence, the projection is modified. In particular, the perpendicular component is a face-on projection and suffers less aliasing. Thus the higher oscillations are damped roughly as $\ell/k\Delta\eta$.

Now all that remains is to evaluate the $q_\perp$ term. Since the Vishniac effect is second order, the convolution theorem for Fourier transforms tells us that

$$\mathbf{q}_\perp = \left( \mathbf{I} - \frac{\mathbf{kk}}{k^2} \right) \frac{1}{2} \sum_{\mathbf{k}'} \mathbf{v}_b(\mathbf{k}') \Delta_b(|\mathbf{k} - \mathbf{k}'|) \qquad (17)$$
$$+ \mathbf{v}_b(\mathbf{k} - \mathbf{k}') \Delta_b(k').$$

We can relate the linear theory baryon velocity and density by the continuity equation (8). Taking the ensemble average of the fluctuation and assuming random phases for the underlying linear theory perturbations, we obtain

$$\langle q_\perp^*(k,\eta) q_\perp(k,\eta') \rangle = \frac{d^2}{2} \dot{D}(\eta) D(\eta) \dot{D}(\eta') D(\eta') \qquad (18)$$
$$\times \sum_{\mathbf{k}'} P(k') P(|\mathbf{k} - \mathbf{k}'|),$$

where $P(k) = |\Delta_b^2(k,\eta_0)|^2$ and the projected vector

$$\mathbf{d} \equiv \left( \mathbf{I} - \frac{\mathbf{kk}}{k^2} \right) \left[ \frac{\mathbf{k}'}{k^2} + \frac{\mathbf{k} - \mathbf{k}'}{|\mathbf{k} - \mathbf{k}'|^2} \right]. \qquad (19)$$

A bit of straightforward but tedious algebra yields

$$\left\langle |\Theta_\ell(k,\eta_0)|^2 \right\rangle = \frac{1}{(4\pi)^2} \frac{1}{\eta_0^2} \frac{\ell(\ell+1)}{k\eta_0} M(k) I_\ell^2(k) P^2(k), \quad (20)$$

where the mode coupling integral, coming from $|\mathbf{d}|^2$, is

$$M(k) = \int_0^\infty dy \int_{-1}^1 dx \frac{(1-x^2)(1-2yx)^2}{(1+y^2-2yx)^2} \qquad (21)$$
$$\times \frac{P[k(1+y^2-2yx)^{1/2}]}{P(k)} \frac{P(ky)}{P(k)}.$$

The integration variables arise from $y = k'/k$ and $\mathbf{k}' \cdot \mathbf{k} = k'kx$ and the time integral is

$$I_\ell(k) = \int_0^{\eta_0} \frac{d\eta}{\eta_0} S_V(\eta) j_\ell(k\Delta\eta). \qquad (22)$$

Here $S_V(\eta) = \dot{D} D \dot{\tau} e^{-\tau} \eta_0^3/(\eta_0 - \eta)$, which reduces to $2(\eta/\eta_0)^3 \dot{\tau} e^{-\tau} \eta_0/(1-\eta/\eta_0)$ if $\Omega_0 = 1$. Note that the integrand is more strongly weighted toward later times than the visibility function itself. It has often been assumed (e.g. Efstathiou 1988) that the contributions come mainly from last scattering where the optical depth $\tau \approx 1$. This causes a significant error in the $k$ to $\ell$ projection, i.e. a given spatial scale would erroneously be considered to contribute to a smaller angle. It can be a very severe effect if the power spectrum peaks toward small scales as in the case of baryon isocurvature scenarios (Peebles 1987). Even extremely small scales can contribute to observable anisotropies at sufficiently late times.

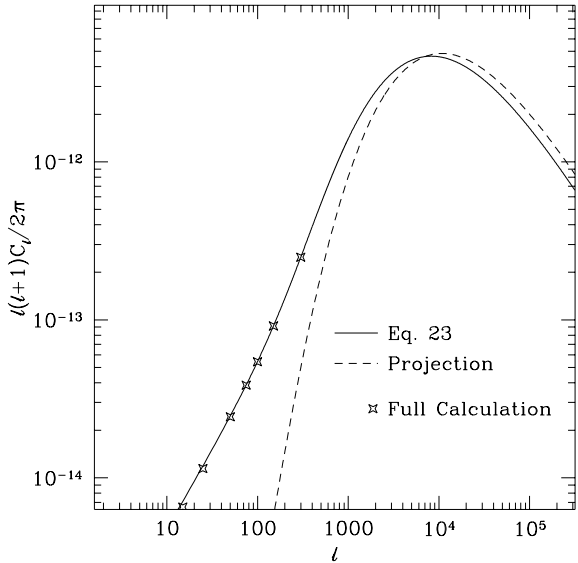

**Fig. 4.** Comparison of different approximations for evaluating the Vishniac effect for a standard CDM model ($\Omega_0 = 1$, $h = 0.5$, $\Omega_b = 0.05$) normalized to COBE ($Q_{\rm rms-PS} = 20\mu$K). The cancellation approximation of equation (23) matches the full integral of equation (22) extremely well, whereas the projection of the r.m.s. spatial fluctuation (Efstathiou 1988) does not.

The time integral in equation (22) is cumbersome to evaluate for high $\ell$ due to the spherical Bessel function. However just as in the first order case, we can exploit the fact that in the cancellation regime the integral extends over many wavelengths of the perturbation. Indeed, we have just shown that the function $S_V$ multiplying $j_\ell$ is even more slowly varying than the wide visibility function itself. Taking $S_V$ out of the integral and employing the approximation of equation (6), we obtain

$$I_\ell(k) \approx \frac{\sqrt{\pi}}{2}\frac{\Gamma[(\ell+1)/2]}{\Gamma[(\ell+2)/2]}\frac{1}{k\eta_0}S_V(\eta_\ell) \qquad (23)$$
$$\approx \sqrt{\frac{\pi}{2\ell}}\frac{1}{k\eta_0}S_V(\eta_\ell), \qquad \ell \gg 1$$

where recall $\eta_\ell = \eta_0 - (\ell + 1/2)/k$. This represents an excellent approximation for angular scales relevant to the Vishniac effect (see Fig. 4).

The random phase assumption for the underlying linear perturbations assures us that there are no cross terms between first and second order contributions or different $k$ modes. Thus the total anisotropy is obtained by integrating over all $k$ modes:

$$C_\ell^{(2)} = \frac{\ell(\ell+1)}{(2\pi)^3}\frac{1}{\eta_0^6}\int \frac{dk}{k}(k\eta_0)^2 M(k) I_\ell^2(k) P^2(k). \qquad (24)$$

In Fig. 5, we plot the Vishniac effect for standard CDM with late to minimal reionization. Since the Vishniac effect is so strongly weighted toward late times, the underlying approximation that Compton drag is negligible holds even for low $\Omega_b h^2$ models unlike for the Doppler effect. The small scale matter power spectrum is well described by an analytic fit (Bond & Efstathiou 1984)

$$P(k) = \frac{Ak}{\left\{1 + [(ak + (bk)^{3/2} + (ck)^2]^\nu\right\}^{2/\nu}}, \qquad (25)$$

where $a = 6.4(\Omega_0 h^2)^{-1}$Mpc, $b = 3.0(\Omega_0 h^2)^{-1}$ Mpc, $c = 1.7(\Omega_0 h^2)^{-1}$ Mpc, and $\nu = 1.13$. The normalization is related to the CMB ensemble averaged quadrupole by $A = 7.62 \times 10^5 (Q_{\rm rms-PS}/20\mu{\rm K})^2 (h^{-1}{\rm Mpc})^4$ in an $\Omega_0 = 1$ model and by employing equation (10) for a $\Lambda$ model. Note that $C_2 = (4\pi/5)(Q_{\rm rms-PS}/2.726{\rm K})^2$.

Most previous calculations (Efstathiou 1988; Hu et al. 1994; Chiba, Sugiyama, & Suto 1994; Persi 1995) of the Vishniac effect have relied upon the approximations of Efstathiou & Bond (1987). This involves approximating the r.m.s. temperature fluctuation at present,

$$|\Theta(k, \eta_0)|^2_{\rm rms} \approx \frac{1}{\eta_0^3}\frac{P^2(k)}{16\pi}M(k)\int_0^{\eta_0}\frac{d\eta}{\eta_0}(1 - \frac{\eta}{\eta_0})^2 S_V^2 \qquad (26)$$

and projecting such fluctuations on the sky today via a free streaming approximation such as

$$\frac{\ell(\ell+1)}{2\pi}C_\ell^{(2)} \approx \frac{1}{2\pi^2}k^3|\Theta(k, \eta_0)|^2_{\rm rms}\bigg|_{k=\ell/(\eta_0-\eta_*)} \qquad (27)$$

where $\eta_*$ is the epoch where optical depth reaches unity. This approach is flawed since the $k$-space power spectrum at the present does not contain sufficient information to reconstruct the anisotropy in $\ell$. As we have seen, anisotropy contributions to a given $\ell$ are widely distributed along the line of sight from last scattering to the present in a form that depends sensitively on the power spectrum. In Fig. 4, we show that this prior approximation causes a $\sim 30\%$ underestimate of the coherence angle of the Vishniac effect in the standard CDM model. The error can be significantly more severe in cases where there is more power at small scales. Our techniques are also more powerful in that they apply to visibility functions with arbitrary features, *e.g.* the bimodal function which occurs in late reionization scenarios.

## 4. Conclusions

We have presented a formalism for calculating CMB anisotropies in the weak coupling or cancellation damped regime. When applied to the Doppler effect in reionized scenarios, it allows for simple post-processing of the standard recombination results. It also gives the $\Lambda$ ISW effect at large angles to good accuracy. Applied to the Vishniac effect, this yields a significantly more accurate description of the Vishniac effect than that which has previously appeared in the literature. As an example of our method we have evaluated the contribution from standard CDM for a range of reionization histories and shown that the arcminute scale anisotropies are small but present even in a minimally ionized scenario.

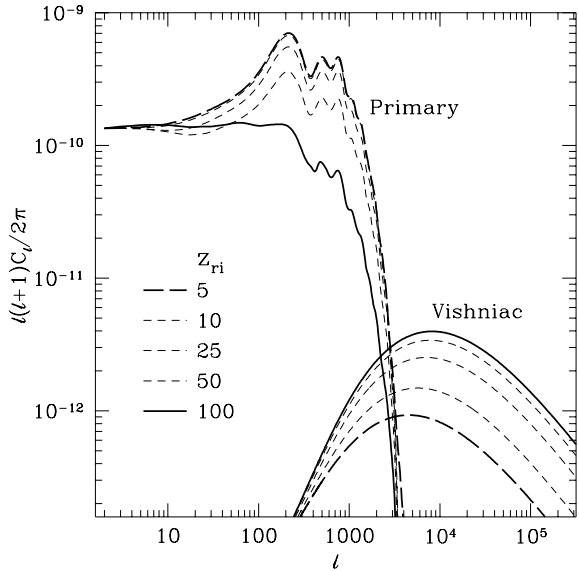

**Fig. 5.** Primary and Vishniac contriubtions for a range of possible ionization histories in the standard CDM model (see Fig. 4). Note that even for minimally ionized $z_{ri} = 5-10$, where first order anisotropies are nearly indistinguishable from the standard recombination case, the Vishniac effect contributes a significant fraction of its total in temperature fluctuations. Because standard CDM has more small scale power than measurements suggest, we expect these calculations to be an upper limit for CDM-like models.

While the small amplitude of arcminute fluctuations in standard flat CDM models will make their detection difficult in the near future, the fact that they are so small implies that their contribution cannot be confused with the much larger effect due to the geometry of the universe. A positive detection of large CMB fluctuations on small scales would be evidence that the geometry of the universe is hyperbolic. As we have shown here the second order contributions in models such as CDM (which already have more small scale power than observationally required) are small. Recently, the non-linear analogue of the Vishniac effect has been calculated from hydrodynamic simulations and shown to be small in comparison with the second order effect in CDM (Persi et al. 1995). Other higher order contributions have also been calculated from simulations. Gravitational lensing only redistributes the power, it does not generate more power on small scales (Seljak 1995a). It has recently been shown (Seljak 1995b) that the Rees-Sciama effect is also small.

Clusters can also induce anisotropies on the CMB from Compton scattering off electrons in the hot cluster medium. Hot electrons transfer energy to the microwave background leading to spectral distortions in the CMB by the Sunyaev-Zel'dovich mechanism. Thus the temperature fluctuation will not only have an angular but also a frequency dependence unlike other sources of anisotropies. Ceballos & Barcons (1994) employ an empirically based model for clusters. They find that in the Rayleigh-Jeans regime, where the Sunyaev-Zel'dovich effect leads to a constant brightness decrement, the anisotropy at arcminutes is on the order $\Delta T/T \lesssim 10^{-7}$. Moreover, the signal is in large part due to bright and easily identifiable clusters. If such known clusters are removed from the sample, the anisotropy drops to an entirely negligible level. One therefore expects that the small Vishniac effect dominates the small angle anisotropy beyond the damping tail of primary anisotropies. Thus the presence of significant arcminute scale anisotropies may provide a robust indicator of curvature in the universe.

## Acknowledgements

We would like to thank D. Scott, J. Silk, and N. Sugiyama for useful comments and discussions. This work was supported in part by grants from the NSF and DOE.